\documentclass[reprint,showpacs,aps,amsmath,amssymb,prd,nofootinbib,floatfix]{revtex4-2}
\usepackage{amsmath,amssymb,graphicx,color}
\usepackage{bm}
\usepackage{epsf}
\usepackage{ulem} 
\usepackage{float}
\usepackage{xcolor}
\usepackage[colorlinks=true]{hyperref}
\hypersetup{citecolor=cyan,linkcolor=magenta}
\usepackage{journals}

\usepackage{mathptmx}
\usepackage{mathrsfs}

\newcommand{\beq}{\begin{equation}}
\newcommand{\eeq}{\end{equation}}
\newcommand{\beqn}{\begin{eqnarray}}
\newcommand{\eeqn}{\end{eqnarray}}
\newcommand{\pa}{\partial}

\newcommand{\varep}{\varepsilon}

\def\cL{\mathscr{L}}

\def\2pi{2\pi}

\begin{document}

\title{Numerical-relativity simulation for tidal disruption of white
  dwarfs by a supermassive black hole}

\author{Alan Tsz-Lok Lam$^1$}
\author{Masaru Shibata$^{1,2}$}
\author{Kenta Kiuchi$^{1,2}$}
\affiliation{$^1$Max Planck Institute for
  Gravitational Physics (Albert Einstein Institute), Am M{\"u}hlenberg 1,
  Potsdam-Golm 14476, Germany}
\affiliation{$^2$Center for Gravitational and Quantum-Information 
  Physics, Yukawa Institute for Theoretical Physics, Kyoto University,
  Kyoto, 606-8502, Japan}

\date{\today}

\begin{abstract}
  We study tidal disruption of white dwarfs in elliptic orbits with the eccentricity of $\sim 1/3$--$2/3$ by a nonspinning supermassive black hole of mass $M_{\rm BH}=10^5M_\odot$ in fully general relativistic simulations targeting the extreme mass-ratio inspiral leading eventually to tidal disruption. Numerical-relativity simulations are performed by employing a suitable formulation in which the weak self-gravity of white dwarfs is accurately solved. We reconfirm that tidal disruption occurs for white dwarfs of the typical mass of $\sim 0.6M_\odot$ and radius $\approx 1.2 \times 10^4$\,km near the marginally bound orbit around a nonspinning black hole with $M_{\rm BH}\alt 4\times 10^5M_\odot$.
\end{abstract}

\pacs{04.25.D-, 04.30.-w, 04.40.Dg}
\maketitle

\section{Introduction}

Tidal disruption of ordinary stars and/or white dwarfs by supermassive black holes has been revealed to be one of the major sources of bright electromagnetic transients (see, e.g., Refs.~\cite{review1,review2,review3}), which have been actively observed in the last decade.  In addition, gravitational waves emitted by tidal disruption of white dwarfs closely orbiting supermassive black holes could be observable by Laser Interferometer Space Antenna (LISA)~\cite{LISA}. 
Electromagnetic signals associated with tidal excitation (e.g., Ref.~\cite{Vick2017}) or mass stripping (e.g., Refs.~\cite{Metzger:2021zia,2022MNRAS.515.4344K,Lu:2022hxq,Linial:2022bjg} for related works) or tidal disruption (e.g., Refs.~\cite{Rosswog1,Rosswog2}) of white dwarfs can be an important electromagnetic counterpart of gravitational waves. Because the expected event rate is not so high~\cite{Macleod} that the signal-to-noise ratio of gravitational waves for the LISA sensitivity is unlikely to be very high, the discovery of the possible electromagnetic counterparts will help extract gravitational waves from the noisy data in the LISA mission.


The condition for mass shedding and tidal disruption during the cross encounter of stars with supermassive black holes is often described by the so-called $\beta$-parameter defined by
\beq
\beta:=\frac{r_t}{r_p},
\eeq
where $r_p$ is the periastron radius for the orbit and $r_t$ is the Hill's radius~\cite{Hill} defined by
\beq
r_t:=R_* \left(\frac{M_{\rm BH}}{M_*}\right)^{1/3},
\eeq
with $R_*$ the stellar radius, $M_*$ the stellar mass, and $M_{\rm BH}$ the mass of the supermassive black hole, respectively. Since the early 1990s (see, e.g., Refs.~\cite{Laguna,Khokhlov93,Diener97}), a large number of numerical simulations have been performed in the last three decades (see, e.g., Refs.~\cite{review4,review5} for reviews of the latest works and Refs.~\cite{GR2019,Ryu1,Ryu2,Ryu3,Ryu4,Ryu5} for some of the most advanced works). They have shown that mass stripping can take place at the close encounter if $\beta$ is larger than about 0.5, and tidal disruption can take place if $\beta \agt 1$ for stars in parabolic orbits (see, e.g., Refs.~\cite{review4,GR2013,Mainetti} for Newtonian simulation works, and also early semianalytical work~\cite{Luminet1,Luminet2}). It is also shown that for close orbits around a black hole, the general relativistic effect can significantly reduce the critical value of $\beta$ for the tidal disruption~\cite{Ryu4}.  Indeed, general relativistic works show that for circular orbits near the innermost stable circular orbit of black holes, the mass shedding can occur even for $\beta \sim 0.4$~\cite{Fishbone,Ishii}. 

However, the previous analyses have been carried out in Newtonian gravity or in relativistic gravity of a black hole with Newtonian (or no) gravity for the companion star or in a tidal approximation with a relativistic tidal potential~\cite{Fishbone,Marck,Ishii}. To date, no fully general relativistic (the so-called numerical-relativity) simulation, i.e., a simulation with no approximation except for the finite differencing, has been done for the tidal disruption problem with $\beta \alt 1$ (but see Refs.~\cite{Haas2012,EP13,East} for a head-on and an off-axis collision).

Numerical-relativity simulation is suitable for the tidal disruption problem for the case that the orbit at the tidal disruption is highly general-relativistic. This is particularly the case for tidal disruption of white dwarfs by supermassive black holes because it can occur only for orbits very close to the black-hole horizon. Advantages of the numerical-relativity simulation are: (i) the redistribution of the energy and angular momentum of the star can be followed in a straightforward manner and (ii) we can directly follow the matter motion after the tidal disruption including the subsequent disk  formation.

In this paper, we present a result of numerical-relativity simulations for tidal disruption of white dwarfs of typical mass ($0.6$--$0.8M_\odot$) by a supermassive black hole with relatively low mass ($M_{\rm BH}=10^5M_\odot$) for the first time. For simplicity, the white dwarfs are modeled by the $\Gamma=5/3$ polytropic equation of state. As a first step toward more detailed and systematic studies, we focus on tidal disruption of white dwarfs in mildly elliptic orbits aiming at confirming that our numerical-relativity approach is suitable for reproducing the criteria of tidal disruption, which has been already investigated in many previous works referred to above.

The paper is organized as follows. In Sec.~\ref{secII}, we describe our formulation for evolving gravitational fields, matter fields, and for providing initial data of a star in elliptic orbits around supermassive black holes. In Sec.~\ref{secIII}, numerical results are presented paying particular attention to the criterion for tidal disruption. Section \ref{secIV} is devoted to a summary.  Throughout this paper, we use the geometrical units of $c=1=G$ where $c$ and $G$ are the speed of light and gravitational constant, respectively. The Latin and Greek indices denote the space and spacetime components, respectively.

\section{Basic equations for the time evolution}\label{secII}

\subsection{Gravitational field}\label{secIIa}

First, we reformulate the Baumgarte-Shapiro-Shibata-Nakamura (BSSN) formalism \cite{shibata1995a,baumgarte1999} in numerical relativity to a form suitable for the simulation of high-mass ratio binaries, in particular for accurately computing a weak self-gravity of white dwarfs. Throughout this paper,  high-mass ratio binaries imply those composed of a very massive black hole of mass $M_{\rm BH} \agt 10^5M_\odot$ and a white dwarf (or an ordinary star) of mass of $M_*=O(M_\odot)$ with the radius $R_* \agt 10^3$\,km, for which the compactness defined by $M_*/R_*$ is smaller than $10^{-3}$.

We consider the two-body problem with a compact orbit of the orbital separation $r \alt 30M_{\rm BH}$. With such setting, the magnitude of the gravitational field generated by the black hole, which is defined by $g_{\mu\nu}-\eta_{\mu\nu}$ is, of order $M_{\rm BH}/r > 10^{-2}$.  Here $g_{\mu\nu}$ and $\eta_{\mu\nu}$ are the spacetime metric and Minkowski metric, respectively. On the other hand, the magnitude of the gravitational field generated by white dwarfs and ordinary stars is of order $M_*/R_* < 10^{-3}$, which is much smaller than that by the black hole. To accurately preserve the nearly equilibrium state of such stars during their orbits, an accurate computation of the gravitational field by them is required. However, if we simply solve Einstein's equation, a numerical error for the computation of the black-hole gravitational field can significantly affect the gravitational field for the white dwarfs/ordinary stars. To avoid this numerical problem, we separate out the gravitational field into the black hole part and other part, although we still solve fully nonlinear equations. The idea employed here is similar to that of Ref.~\cite{EP13}, but we develop a formalism based on the BSSN formalism.

In a version of the BSSN formalism \cite{baumgarte1999}, the basic equations are written in the form: 
\beqn 
&&(\pa_{t} - \beta^k \pa_k) \tilde
\gamma_{ij} =-2\alpha \tilde A_{ij} +\tilde \gamma_{ik} \pa_j
\beta^k+\tilde \gamma_{jk} \pa_i \beta^k
-\frac{2}{3}\tilde \gamma_{ij} \pa_k \beta^k,~~~~~ \\
\label{heq:bssn}
&&(\pa_{t} - \beta^l \pa_l) \tilde A_{ij} 
= W^2\biggl[ \alpha \left(R_{ij}
-\frac{\gamma_{ij}}{3} R_k^{~k} \right) \nonumber \\
&&\hskip 2cm -\left( D_i D_j \alpha - \frac{\gamma_{ij}}{3}D_k D^k \alpha \right)
\biggr] \nonumber \\
&& \hskip 2.cm
+\alpha \left(K \tilde A_{ij} - 2 \tilde A_{ik} \tilde A_j^{~k}\right)
\nonumber \\
&& \hskip 2.cm
+ \tilde A_{kj} \pa_i \beta^k 
+\tilde A_{ki} \pa_j \beta^k 
-\frac{2}{3} \tilde A_{ij} \pa_k \beta^k \nonumber \\
&& \hskip 2.cm
-8\pi \frac{G}{c^4} \alpha W^2 \left( 
S_{ij}-\frac{1}{3} \gamma_{ij} S_k^{~k}
\right), \label{aijeq:bssn} \\
&&(\pa_{t} - \beta^l \pa_l) W = \frac{W}{3}\left( 
\alpha K - \pa_k \beta^k \right), \label{peq:bssn} 
\eeqn
\beqn
&&(\pa_{t} - \beta^l \pa_l) K 
=\alpha \left[ \tilde A_{ij} \tilde A^{ij}+\frac{1}{3}K^2 \right]
\nonumber \\
&&\hskip 2.2cm -W^2 \left(
\tilde D_k \tilde D^k \alpha
-\frac{\pa_i W}{W}\tilde \gamma^{ij} \pa_j \alpha
\right)\nonumber \\
&&\hskip 2.2cm +4\pi \frac{G}{c^4}\alpha
\left(\rho_{\rm h}+ S_k^{~k}\right), \label{keq:bssn}
\eeqn
\beqn
&&(\pa_{t} - \beta^l \pa_l) \tilde \Gamma^i 
=-2\tilde{A}^{ij} \partial_j\alpha
+2\alpha\left[
\tilde{\Gamma}^{i}_{~jk}\tilde{A}^{jk} \right.
\nonumber \\
&& \hskip 1.5cm \left.
-\frac{2}{3}\tilde{\gamma}^{ij} \pa_j K
-8\pi \frac{G}{c^4}\tilde{\gamma}^{ik} J_k
-3\frac{\pa_j W}{W}\tilde{A}^{ij}
\right] \nonumber \\
&&\hskip 1.2cm -\tilde{\Gamma}^j\partial_j\beta^i
+\frac{2}{3}\tilde{\Gamma}^i\partial_j\beta^j
+\frac{1}{3}\tilde{\gamma}^{ik}\pa_k\pa_{j}\beta^j
+\tilde{\gamma}^{jk} \pa_{j}\pa_k\beta^i, \nonumber \\
\label{eq:Gamma} 
\eeqn
where $\alpha$ is the lapse function, $\beta^j$ is the shift vector, $\tilde \gamma_{ij}$ is the conformal three metric defined from the three metric $\gamma_{ij}$ by $\tilde \gamma_{ij}:=\gamma^{-1/3}\gamma_{ij}$ with $\gamma={\rm det}(\gamma_{ij})$, $W:=\gamma^{-1/6}$, $\tilde A_{ij}$ is the conformal trace-free extrinsic curvature defined from the extrinsic curvature $K_{ij}$ by $\tilde A_{ij}=W^2(K_{ij}-\gamma_{ij}K/3)$ with $K:=K_k^{~k}$, and $\tilde \Gamma^i:=-\pa_j \tilde \gamma^{ij}$. $\rho_{\rm h}$, $J_i$, and $S_{ij}$ are quantities defined from the energy-momentum tensor, $T_{\mu\nu}$, by $\rho_{\rm h}=T^{\mu\nu} n_\mu n_\nu$, $J_i=-T^{\mu\nu} n_\mu \gamma_{\nu i}$, and $S_{ij}=T^{\mu\nu}\gamma_{\mu i}\gamma_{\nu j}$ with $n^\mu$ the timelike unit vector normal to spatial hypersurfaces.

In this problem, we employ the so-called puncture gauge~\cite{Alcubierre:2002kk}, in which the evolution equations for $\alpha$ and $\beta^i$ are written as
\beqn
\pa_t \alpha &=&  - 2\alpha K,\label{eq:alpha} \\
\pa_{t} \beta^i &=& \frac{3}{4} B^i,\\
\pa_{t} B^i &=&
\pa_{t} \tilde\Gamma^i -\eta_B B^i, \label{eq:gammai}
\eeqn
where $B^i$ is an auxiliary three-component variable and $\eta_B$ is a constant of order $M_{\rm BH}^{-1}$. 

By introducing a static black-hole solution for the geometric variables, $\alpha_{0}$, $\beta_0^i$, $\tilde \gamma^0_{ij}$, $W_0$, $\tilde A^0_{ij}$, and $K_0$ and by writing all the variables by
\beqn
\alpha&=&\alpha_0 + \alpha_{\rm s},\\
\beta^i&=&\beta_0^i + \beta_{\rm s}^i,\\
\tilde \gamma_{ij}&=&\tilde \gamma^0_{ij} + \tilde \gamma^{\rm s}_{ij},\\
W&=&W_0 + W_{\rm s},\\
\tilde A_{ij}&=&\tilde A^0_{ij} + \tilde A^{\rm s}_{ij},\\
K&=&K_0 + K_{\rm s},\\
\tilde \Gamma^{i}&=&\tilde \Gamma_0^i + \tilde \Gamma_{{\rm s}}^i,
\eeqn
we then write down the equations for $\alpha_{\rm s}$, $\beta_{\rm s}^i$, $\tilde \gamma^{\rm s}_{ij}$, $W_{\rm s}$, $\tilde A^{\rm s}_{ij}$, $K_{\rm s}$, and $\tilde \Gamma_{\rm s}^i$ (these are denoted by a representative variable $Q_s$ as follows). Specifically, the evolution equations (\ref{heq:bssn})--(\ref{eq:Gamma}) and (\ref{eq:alpha})--(\ref{eq:gammai}) 
of the geometrical variables (denoted by a representative variable $Q$) are schematically written in the form 
\beq
\pa_t Q= F(Q).
\eeq
Then, for the decomposition of $Q=Q_0 + Q_{\rm s}$ with $F(Q_0)=0$ (under the conditions of $\pa_t Q_0=0$), we write the equation for $Q_{\rm s}$ as
\beq
\pa_t Q_{\rm s}= F(Q_0+Q_{\rm s})-F(Q_0).
\eeq
In numerical simulation, $F(Q_0)$ obtained from finite difference is nonzero which contains the truncation error of evolving the stationary background metric numerically.
Here, we added the second-term in the right-hand side to explicitly subtract the leading error of evolving the background metric so that the right-hand side of the evolution equation of $Q_{\rm s}$ does not have the zeroth order terms in $Q_{\rm s}$

Any static black-hole solutions can be used for $\alpha_0$, $\beta_0^i, \cdots$, but in the BSSN formalism with the puncture gauge, the metric relaxes to a solution in the limit hypersurface with $K_0=0$. 
Using such a trumpet-puncture black hole also allows us to construct the initial data in the conformal-thin-sandwich (CTS) formalism~\cite{Baumgarte2011} (see Sec.~\ref{secIIc}).
Thus, in the present formalism, it is appropriate to employ such a solution. In the nonspinning black hole, the analytic solution is known and is written as~\cite{E73}
\beqn
\alpha_0&=&\sqrt{1 -\frac{2M_{\rm BH}}{R}+\frac{27 M_{\rm BH}^4}{16 R^4}},\\
\beta_0^i&=&\frac{3\sqrt{3} M_{\rm BH}^2}{4R^3}x^i,\\
W_0&=&\frac{r}{R},\\
\tilde \gamma^0_{ij}&=&\delta_{ij},~~{\rm i.e.},~~\tilde \Gamma_0^i=0, \\
\tilde A^0_{ij}&=&\frac{3\sqrt{3}M_{\rm BH}^2}{4R^3}\left(\delta_{ij}
-3 \frac{x^i x^j}{r^2}\right),
\eeqn
and $K_0=0$ where $R$ is a function of $r$ determined by~\cite{BN2007}
\beqn
r&=&\left(\frac{2R + M_{\rm BH} + \sqrt{4R^2 + 4M_{\rm BH}R + 3M_{\rm BH}^2}}{4}\right)
\nonumber \\
&& \times
\left[\frac{(4+3\sqrt{2})(2R-3M_{\rm BH})}
    {8R+6M_{\rm BH}+3\sqrt{8R^2+8M_{\rm BH}R+6M_{\rm BH}^2}}\right]^{1/\sqrt{2}}. 
\eeqn
We note that $r=0$ corresponds to $R=3M_{\rm BH}/2$ and the event horizon is located at $R=2M_{\rm BH}$ (i.e., $r \approx 0.78M_{\rm BH}$) in this solution. 

\subsection{Hydrodynamics}

In this paper we model white dwarfs simply by the polytropic equation of state,
\beq
P=\kappa \rho^\Gamma,
\eeq
where $P$ and $\rho$ are the pressure and rest-mass density, respectively, $\kappa$ the polytropic constant, and $\Gamma$ adiabatic index for which we set to be $5/3$. 
For the hydrodynamics, we solve the continuity and Euler equations,
\beqn
\nabla_\mu (\rho u^\mu)=0,\label{continuity} \\
\nabla_\mu T^\mu_{~k}=0, \label{Euler}
\eeqn
with $\nabla_\mu$ the covariant derivative with respect to $g_{\mu\nu}$
and 
\beq
T_{\mu\nu}=(\rho + \rho \varep + P)u_\mu u_\nu  + P g_{\mu\nu},
\eeq
where $\varep$ and $u^\mu$ are the specific internal energy and four velocity, respectively. In this work we do not solve the energy equation, and determine $\varep$ simply by $\varep=\kappa\rho^{\Gamma-1}/(\Gamma-1)$ which is derived from the condition that the specific entropy is conserved for the fluid elements. The continuity and Euler equations are solved in the same scheme as that used in Refs.~\cite{SACRA,SACRAMP}.

The motivation for using the polytropic equation of state comes from the fact that our primary purpose of this paper is to explore the tidal disruption condition for a relatively low value of $\beta < 1$ and the formation of shocks by the tidal compression does not play any role. We here focus only on the process of tidal disruption and subsequent short-term evolution of the tidally disrupted material. After the tidal disruption, the fluid is highly elongated and during the long-term evolution of the fluid elements with different specific energy and angular momentum, they collide and shocks are likely to be formed. For such a phase, the shock heating will play an important role. Our plan is to follow this phase by solving the energy equation with a more general equation of state.

\subsection{Initial condition}\label{secIIc}

First, we describe the formulation employed in this paper for computing the initial data in which white dwarfs are approximately in an equilibrium state in their comoving frame. From Eq.~(\ref{Euler}),
we have
\beqn
\rho u^\mu \nabla_\mu (h u_i) + \nabla_i P=0,\label{eq3}
\eeqn
where $h$ is the specific enthalpy defined by $h:=1+\varep+P/\rho$. To derive Eq.~(\ref{eq3}), we used Eq.~(\ref{continuity}). 

For the isentropic fluid, the first law of thermodynamics is written as
\beq
\rho dh=dP, \label{first}
\eeq
where $dQ$ denotes the variation of a quantity $Q$ in the fluid rest frame. In the polytropic equations of state employed in this work, we obtain the relation
\beqn
h=\int \frac{dP}{\rho} ~~{\rm and}~~ \ln h =\int \frac{dP}{\rho h}. \label{eq6}
\eeqn
In this situation, Eq.~(\ref{eq3}) is rewritten to
\beqn
u^\mu \nabla_\mu (h u_i) +\pa_i h=0. 
\label{eq8}
\eeqn

Then, we define $k^\mu:=u^\mu/u^t$. Using this quantity, Eq.~(\ref{eq8}) is written to
\beq
u^t \cL_k (h u_i) -u^t h u_\mu \nabla_i k^\mu
+\pa_i h=0, \label{eq8a}
\eeq
where $\cL_k$ denotes the Lie derivative with respect to $k^\mu$. The second term of Eq.~(\ref{eq8a}) is written as
\beqn
 u^t h u_\mu \nabla_i k^\mu
= u^t h u_\mu \nabla_i (u^\mu/u^t)= h \pa_i \ln u^t,
\eeqn
where we used $u^\mu u_\mu=-1$. Thus, Eq.~(\ref{eq8a}) is written to
\beq
\cL_k (h u_i)+\pa_i (h/u^t)=0. \label{eq8b}
\eeq

We consider an initial condition for a system composed of a star of mass $M_*$ and radius $R_*$, for which the center is located on the $x$-axis, around a massive black hole of mass $M_{\rm BH} \gg M_*$ and $M_{\rm BH} \gg R_*$ which is located at a coordinate origin. We assume that the star predominantly moves toward the $y$-direction with the identical specific momentum.  Thus we set $v^i:=u^i/u^t=-\beta^i_0+V^i$ where $V^i=V\delta^i_{~y}$ with $V$ being a constant to be determined. Here the term of $\beta^i_0$ is added to simplify the iteration process for computing quasiequilibrium states. Then, $u^t$ is calculated from
\beq
u^t = \left[ \alpha^2- \gamma_{ij}(v^i+\beta^i)(v^j+\beta^j)\right]^{-1/2}. 
\eeq




In the present context, $\cL_k (h u_i)$ can be assumed to be zero for $i=y$ and $z$, because the star has momentarily translation invariance for the motion toward the $y$- and $z$-directions.  By contrast, with respect to the $x$-direction, the star receives the force from the massive black hole. Since the radius of the star, $R_*$, is much smaller than the orbital separation, $x_0$, and $x_0$ is larger than the black-hole radius of $\sim M_{\rm BH}$, $\cL_k (h u_i)$ for the $x$–direction can be approximated by $\pa_i [A(x-x_0)]$ where we take $A$ to be a constant, which should be approximately written as $\sim -M_{\rm BH}/x_0^2$ for $x_0 > 0$. Then, Eq.~(\ref{eq8}) is integrated to give
\beq
A(x-x_0) + \frac{h}{u^t}=C, \label{eq11}
\eeq
where $C$ is an integration constant. We note that Eq.~(\ref{eq11}) is not an exact first integral of the Euler equation but can be considered as an approximate one for obtaining an initial condition in which the star is in an approximate equilibrium state.

For computing initial conditions, we assume the line elements of the form
\beq
ds^2=-(\alpha^2 - \beta_k \beta^k)dt^2 + 2 \beta_k dt dx^k +
\psi^4 \delta_{ij} dx^i dx^j, \eeq
where $\psi$ is the conformal factor. Using the Isenberg-Wilson-Mathews formalism~\cite{IWM1,IWM2}, the basic equations are written as
\beqn
\Delta \psi&=&-2\pi \rho_{\rm H} \psi^5
- \frac{\psi^{-7}}{8} \hat A_{ij} \hat A^{ij},\label{psieq} \\
\Delta (\alpha\psi)&=&2\pi \alpha \psi^5 (\rho_{\rm H}+2S)
+ \frac{7}{8} \alpha\psi^{-7} \hat A_{ij} \hat A^{ij},~~~\label{Xeq} \\
\pa_i \hat A^{i}_{~j}&=&8 \pi J_j \psi^6,\label{aijeq}
\eeqn
where
\beqn
\rho_{\rm H}&=&\rho h (\alpha u^t)^2 - P,\\
S&=&\rho h [(\alpha u^t)^2-1] + 3 P,\\
J_i&=&\rho h \alpha u^t u_i,
\eeqn
and $\Delta$ is the flat Laplacian. $\hat A_{ij}$ is defined from the
extrinsic curvature, $\tilde A_{ij}$, by $\hat A_{ij}=\psi^6\tilde A_{ij}$ 
and $K$ is set to be zero. 
Using the CTS decomposition~\cite{York1999,Baumgarte2007} with trumpet-puncture
\beqn
&&\hat A^{ij}=\hat A^{ij}_{0} +
\frac{\psi^6}{2\alpha}
\left(\mathbf{L} \beta_{\rm s} \right)^{ij},\label{cts}
\eeqn
Equation~(\ref{aijeq}) is rewritten as
\beqn
\pa_j \pa^j \beta^i_{\rm s} + \frac{1}{3}\delta^{ij}\pa_j \pa_k \beta^k_{\rm s}
&=& 16 \pi \alpha \psi^4 J^i \nonumber\\
&+& \left(\mathbf{L} \beta_{\rm s} \right)^{ij} \pa_j \ln\left(\alpha \psi^{-6} \right),\label{beq}
\eeqn
where $\left(\mathbf{L} \beta_{\rm s} \right)^{ij}=
\left(
\delta^{ik} \pa_k \beta^j_{\rm s} + \delta^{jk} \pa_k \beta^i_{\rm s} -\frac{2}{3}\delta^{ij}\pa_k \beta^k_{\rm s}
\right)$.
Note that although there are some works in constructing binary black holes initial data with trumpet-puncture~\cite{Immerman2009,Tim2014,Clemente2017}, this is, to our knowledge, the first attempt combining the CTS decomposition and puncture method with the limit (trumpet)  hypersurface in constructing quasi-equilibrium initial data in nonvacuum spacetime.
We assume that the contribution to the extrinsic curvature from the black hole is negligible because the orbital momentum of the black hole is negligible in this problem, and thus, we set the black hole at rest (however, it is straightforward to take into account the small black-hole motion \cite{Slinker2018} in our formalism.)


For a solution of the initial data, we have to determine the free parameters, $A$, $C$, and $V$. In the polytropic equation of state, we can consider $\kappa$ as well as the central density $\rho_c$ as free parameters.  In the following, we first consider that $V$ and rest mass of the star are input parameters and $A$, $C$, and $\kappa$ are parameters to be determined during the iteration process in numerical computation. Our method to adjust $\kappa$ to a desired value will be described later.

To determine these three parameters we need three conditions, for which we choose the following relations. First, we fix the location of the surface of white dwarfs along the $x$-axis as $x=x_1$ (referred to as point 1) and $x=x_2$ (point 2). Typically, we choose $x_1+x_2=2x_0$.  At the surface, $h=1$, and thus, Eq.~(\ref{eq11}) gives
\beq
A(x_1-x_0)+\frac{1}{u^t_1}=A(x_2-x_0)+\frac{1}{u^t_2}=C,
\label{cond1}
\eeq
where $u^t_1$ and $u^t_2$ are the values of $u^t$ at points 1 and 2.
In addition, we fix the rest mass of the star which is defined by
\beq
m_*=\int d^3x \rho \alpha \psi^6 u^t, \label{eqmm}
\eeq
where $m_*$ is approximately equal to the gravitational mass $M_*$ because the star is only weakly self-gravitating.

Using the condition (\ref{cond1}), the values of $C$ and $A$ are determined, and subsequently, $h$ is determined from Eq.~(\ref{eq11}). In the polytropic equation of state, the rest-mass density is written as
\beq
\rho=\left(\frac{(h-1)(\Gamma-1)}{\kappa \Gamma}\right)^{1/(\Gamma-1)},
\label{eq12}
\eeq
and thus, from Eq.~(\ref{eqmm}), $\kappa$ is determined for given values of $m_*$ and $x_2-x_1$. Once these free parameters are determined, the rest-mass density are obtained from Eq.~(\ref{eq12}).

For realistic setting, we have to obtain the desired values of the mass of the star and the value of $\kappa$. The value of $\kappa$ is controlled by varying the stellar diameter $x_2-x_1$ for a given value of $m_*$. 

To take into account the effect of the black-hole gravity, we employ the puncture formulation by setting
\beqn
\psi&=&\psi_0 + \phi, \\
\alpha\psi&=&\alpha_0\psi_0 + X,\\
\beta^k&=&\beta^k_0 + \beta^k_{\rm s},\\
\hat A_{ij}&=&\hat A_{ij}^0 + \hat A^{\rm s}_{ij}, 
\eeqn
where $\psi_0$, $\alpha_0$, $\beta^k_0$, and $\hat A_{ij}^0$ denote the solutions of vacuum Einstein's equation shown already in Sec.~\ref{secIIa}. Then we numerically solve the equations for $\phi$, $X$, $\beta^k_{\rm s}$, and $\hat A^{\rm s}_{ij}$ from Eqs.~(\ref{psieq}), (\ref{Xeq}), (\ref{beq}), and (\ref{cts}). 
The initial data is prepared using the {\tt octree-mg} code~\cite{octreemg},
an open source multigrid library with an octree adaptive-mesh refinement (AMR) grid, 
which we modified to support a fourth-order finite-difference elliptic solver.

\section{Numerical simulation}\label{secIII}

\subsection{Setup}

The simulation is performed using an AMR algorithm with the equatorial symmetry imposed on the $z=0$ (equatorial) plane using the {\tt SACRA-TD} code (for {\tt SACRA} see Refs.~\cite{SACRA,SACRAMP}). We prepare two sets of finer domains, one of which comoves with a white dwarf and the other of which is located around the center and covers the massive black hole. Because the radius of the white dwarf, $R_*$, is smaller than the black-hole horizon radius $\sim M_{\rm BH}$, we need to prepare more domains for resolving the white dwarf. In addition to these domains, we prepare coarser domains that contain both the finer domains in their inside.  All the domains are covered by $(2N+1, 2N+1, N+1)$ grid points for $(x, y, z)$ with $N$ being an even number.

Specifically, each domain is labeled by $i$ which runs as $0,1,2,\cdots, i_{\rm fix},\cdots,i_{\rm BH},\cdots, i_{\rm max}$. The grid resolution for the domains with $i_{\rm fix} \leq i \leq i_{\rm BH}$ is identical with that with $i_{\rm BH}+1 \leq i \leq 2i_{\rm BH}-i_{\rm fix}+1 (< i_{\rm max})$, respectively. 
For $0 \leq i \leq i_{\rm BH}$, the center of the domain is located at the origin, at which a black hole is present. Strictly speaking, the black hole moves due to the backreaction against the motion of the companion star, but this motion is tiny because of the condition $M_{\rm BH} \gg M_*$. For these domains, the $i$th level covers a half cubic region of $[-L_i:L_i] \times [-L_i:L_i] \times [0:L_i]$ where $L_i=N\Delta x_i$, $\Delta x_i$ is the grid spacing for the $i$th level, and the grid spacing for each level is determined by $\Delta x_{i+1}=\Delta x_{i}/2$ ($i=0,1,2, \cdots , i_{\mathrm{BH}}-1$ and $i=i_{\mathrm{BH}}+1, \cdots, i_{\rm max}-1$) with $\Delta x_{i_{\mathrm{BH}}+1}=\Delta x_{i_{\rm fix}}$ 
and  $L_{i_{\mathrm{BH}}} \sim 0.8 M_{\rm BH}$.

\begin{figure}[t]
\includegraphics[width=86mm]{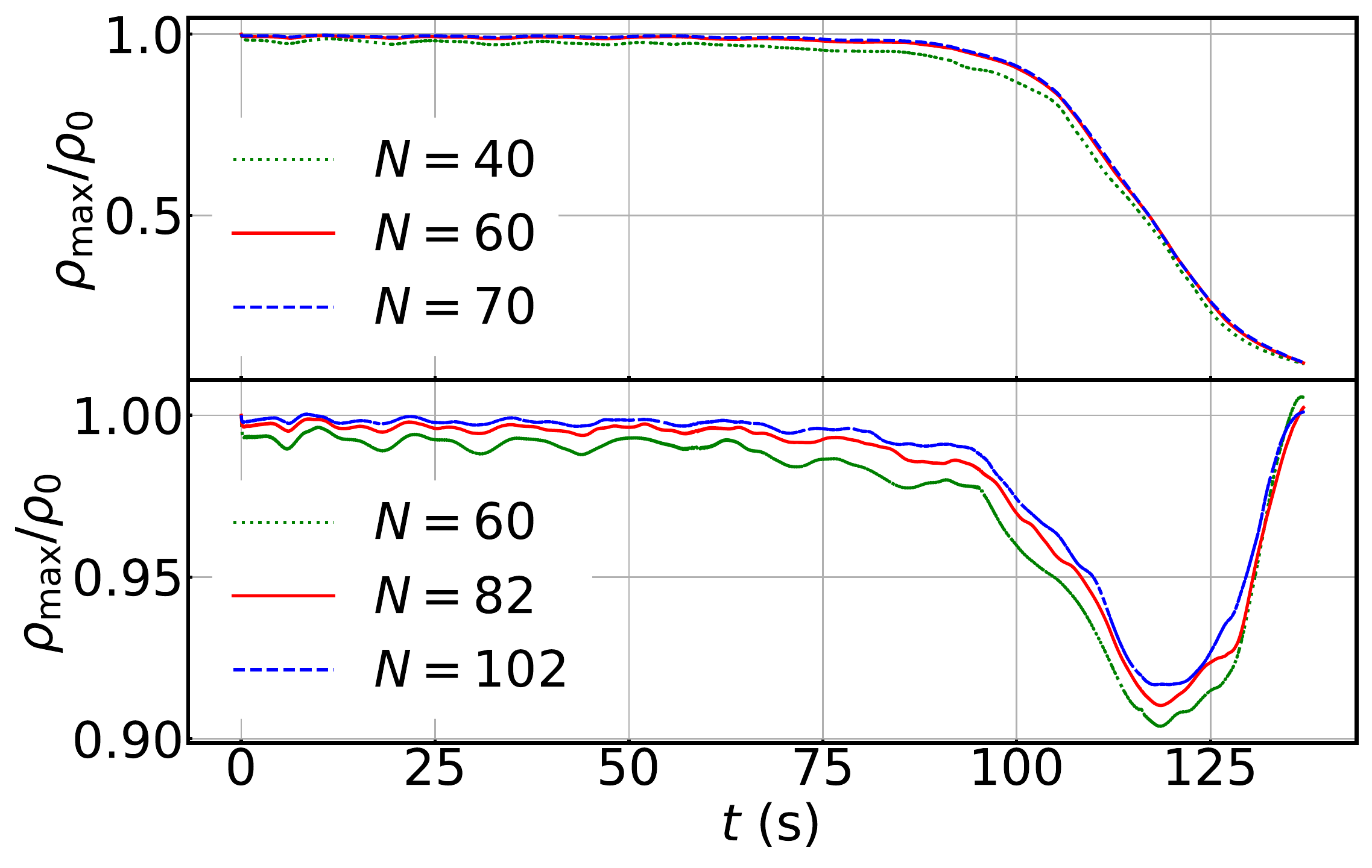}
\caption{Maximum density as a function of time for model \texttt{M7V165} (top) with $N=40$, $60$, $70$, and model \texttt{M8V17} (bottom) with $N=60$, $82$, $102$. We find that a fair convergence is obtained with $N=60$. 
\label{conv}}
\end{figure}

For the moving domains that cover the white dwarf, the center is chosen to approximately agree with the location of the density maximum. In the present context, the local density maximum is approximately located along a geodesic around the supermassive black hole. 
The size of the finest domain with $i=i_{\rm max}$, $L_{\rm max}$, is chosen so that it is $1.3$--$1.5R_*$. 
We check the convergence of two different models with three grid resolutions as illustrated in Fig.~\ref{conv}.
Higher resolution is used for model \texttt{M8V17} to measure the spin up of the white dwarf more accurately (see Sec.~\ref{NumericalResults}).
We obtain good convergence for both models, and thus, we employ $N=60$ as the standard resolution in this paper.

\begin{table}[t]
  \caption{Models considered in this paper and the fate (last column). 
    \texttt{M7V16a} and \texttt{M7V16b} correspond to the models with $R_*=8.5\times 10^3$ and $7.0 \times 10^3$\,km, respectively. For other models, $R_*\approx 10^4 (M_*/0.7M_\odot)^{-1/3}$\,km.
    $r_p$ and $r_{p,A}$ are periastron radius in the present coordinates and the Schwarzschild coordinates, respectively.
    TD and OC denote tidal disruption and appreciable oscillation of white dwarfs,
    and NN denotes that no appreciable tidal effect is found. 
  }
\begin{tabular}{cccccccc} \hline
  ID & ~$V$~ & $M_*(M_\odot)$ & $r_p/M_{\rm BH}$ & $r_{p,A}/M_{\rm BH}$ & $J/M_{\rm BH}$
  & ~~$\beta$~~ & ~~Fate~~ \\
  \hline \hline
  \texttt{M6V16} & 0.160 & 0.6  & 4.401 &  5.456 & 3.775 & 0.72  & TD \\
  \texttt{M7V16} & 0.160 & 0.7  & 4.401 &  5.456 & 3.775 & 0.65  & TD \\
  \texttt{M7V16a} &0.160 & 0.7 & 4.401 &  5.456 & 3.775 & 0.55  & TD \\
  \texttt{M7V16b} &0.160 & 0.7 & 4.401 &  5.456 & 3.775 & 0.45  & TD/OC \\
  \texttt{M8V16} & 0.160 & 0.8  & 4.401 &  5.456 & 3.775 & 0.59  & TD \\
  \texttt{M7V165} & 0.165 & 0.7  & 5.770 &  6.813 & 3.897 & 0.52  & TD \\
  \texttt{M8V165} & 0.165 & 0.8  & 5.770 &  6.813 & 3.897 & 0.47  & TD/OC \\
  \texttt{M6V17} & 0.170 & 0.6  & 7.030 &  8.065 & 4.019 & 0.49  & TD/OC \\
  \texttt{M7V17} & 0.170 & 0.7  & 7.030 &  8.065 & 4.019 & 0.44  & OC \\
  \texttt{M8V17} & 0.170 & 0.8  & 7.030 &  8.065 & 4.019 & 0.40  & OC \\
  \texttt{M6V175} & 0.175 & 0.6  & 8.317 &  9.346 & 4.142 & 0.42  & OC \\
  \texttt{M7V18} & 0.180 & 0.7 & 9.681 & 10.707 & 4.265 & 0.33  & NN \\
 \hline
\end{tabular}
\label{table1}
\end{table}


\subsection{Numerical results}\label{NumericalResults}

\begin{figure}[t]
\includegraphics[width=86mm]{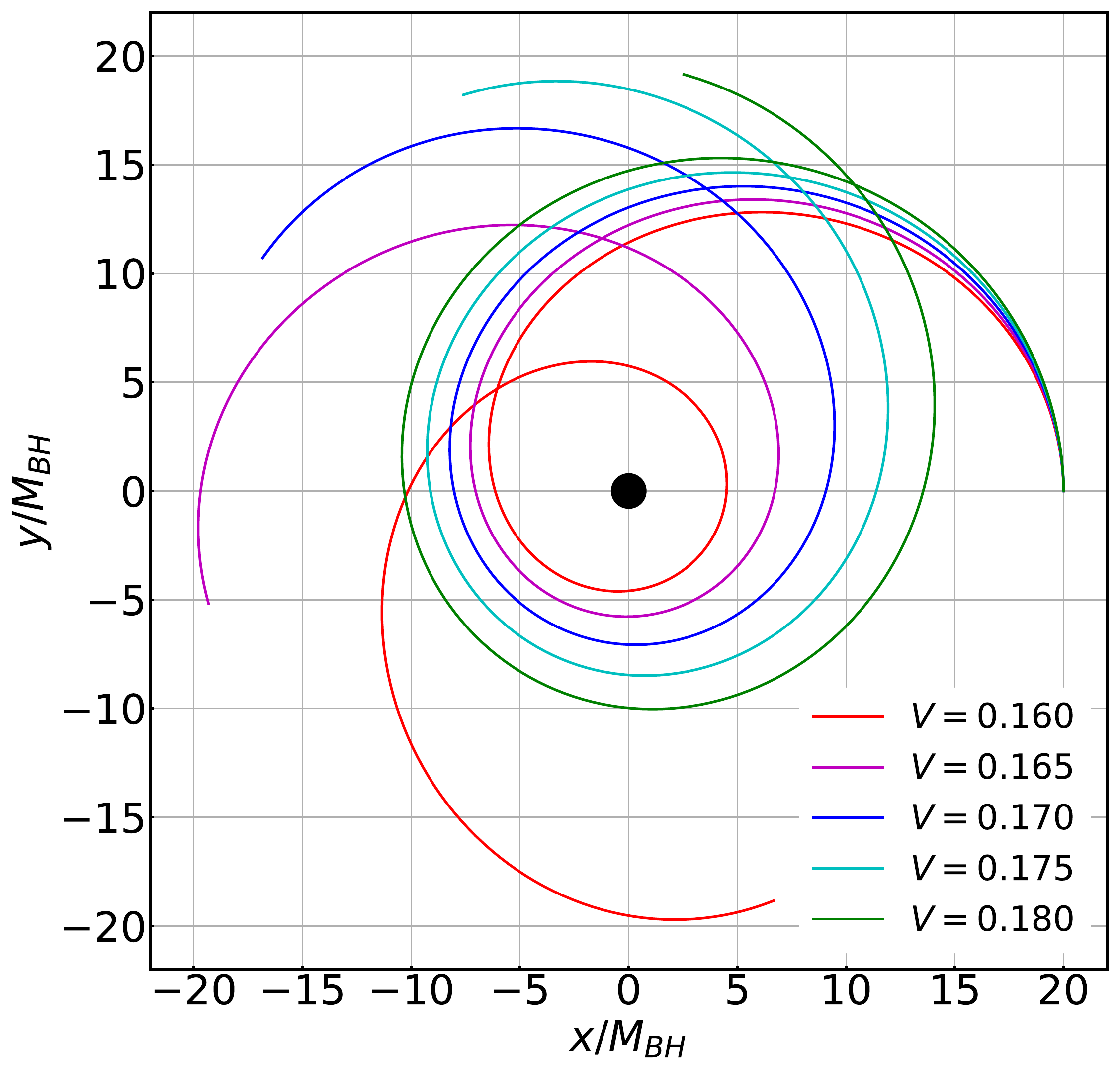}
\caption{Geodesics for $V=0.160$, 0.165, 0.170, 0.175 and 0.180 in the coordinates of $g_{\mu\nu}^0$. Those only for one orbital period started from $x=20M_{\rm BH}$ and $y=z=0$ are plotted. The filled circle at the center represents the black hole with the coordinate radius of its event horizon $r \approx 0.78M_{\rm BH}$. 
\label{orbit}}
\end{figure}

In the present paper we focus on the case that the black-hole mass is $M_{\rm BH}=10^5M_\odot$, the white-dwarf mass is $M_*=0.6$, $0.7$, and $0.8M_\odot$. For the polytropic equation of state, the stellar radius, $R_*$, is proportional to $M_*^{(\Gamma-2)/(3\Gamma-4)}$ for a fixed value of $\kappa$. Thus, for $\Gamma=5/3$, the stellar radius depends only weakly on the stellar mass.  In the present case we basically choose the value of $\kappa$ so that $R_* \approx 1.0\times 10^4 (M_*/0.7M_\odot)^{-1/3}$\,km. For $M_*=0.7 M_\odot$ and $V=0.160$, we also prepare two additional cases where $\kappa$ is chosen such that $R_*=8.5 \times 10^3$ km and $R_*=7.0 \times 10^3$ km.

The initial separation is set to be $x_0=20M_{\rm BH}$ (it is $\approx 21.01M_{\rm BH}$ in the Schwarzschild coordinates), and $V$ is chosen to be $0.160$, $0.165$, $0.170$, $0.175$, and $0.180$ (see Table~\ref{table1}). The corresponding specific angular momentum of the white dwarf is $J\approx 3.7748$, $3.8968$, $4.0192$, $4.142$, and $4.2653 M_{\rm BH}$, and the resulting periastron radius is $r_p/M_{\rm BH} (r_{p,A}/M_{\rm BH})=4.401 (5.456)$, 5.770 (6.813), 7.030 (8.065), 8.317 (9.346), and 9.681 (10.707) where in the parenthesis the values in the Schwarzschild coordinate, i.e., areal radius (hereafter denoted by $r_{p,A}$), are described. In Fig.~\ref{orbit}, we plot the geodesics only for one orbital period for $V=0.160$, 0.165, 0.170, and 0.180.

With these settings, the white dwarf has an elliptic orbit around the black hole with the periastron at $r_p \approx (4.4$--$10)M_{\rm BH}$, and thus, the eccentricity is approximately defined by $e=(x_0-r_p)/(x_0+r_p)$ is $\approx 1/3$--$2/3$. Here, $x_0(=20M_{\rm BH})$ and $r_p$ are defined in the radial coordinates of the metric of $g^0_{\mu\nu}$, and thus, the values of $e$ slightly change if we define it in the areal coordinate (Schwarzschild radial coordinate).

For the models mentioned above, the value of $\beta$ is in the range between 0.33 and 0.72 and estimated by
\beqn
\beta &\approx& 0.59 \left(\frac{R_*}{10^4\,{\rm km}}\right)
\left(\frac{M_*}{0.7M_\odot}\right)^{-1/3} \nonumber \\
&&~~~~ \times \left(\frac{r_{p,A}}{6M_{\rm BH}}\right)^{-1}
\left(\frac{M_{\rm BH}}{10^5M_\odot}\right)^{-2/3}, 
\eeqn
where the areal radius $r_{p,A}$ is used for the definition of $\beta$ in this section.
For $V=0.160$ and $0.165$ with $M_*=0.6$--$0.8M_\odot$, we find $0.50 \leq \beta \leq 0.7$, and thus, the white dwarf is expected to be strongly perturbed by the black-hole tidal field for $M_*=0.6$--$0.8M_\odot$. By contrast, for $V=0.180$, $\beta < 0.35$ with $M_*=0.7M_\odot$, and thus, the tidal force of the black hole is likely to be too weak to perturb the white dwarf. 

For $V=0.170$, $\beta \approx 0.49$, $0.44$, and $0.40$ with $M_*=0.6$, $0.7$, and $0.8M_\odot$ respectively. In these cases, tidal disruption is not very likely to take place but the tidal force from the black hole should induce the stellar oscillation on the white dwarf. Because for $\Gamma=5/3$, the stellar radius depends only weakly on the stellar mass, the presence or absence of the tidal disruption is likely to depend primarily on the value of $V$ (or the specific angular momentum of the white dwarfs) in the present setting. In the following, we will show that our code can reproduce all these expected phenomena.

\begin{figure}[t]
\includegraphics[width=88mm]{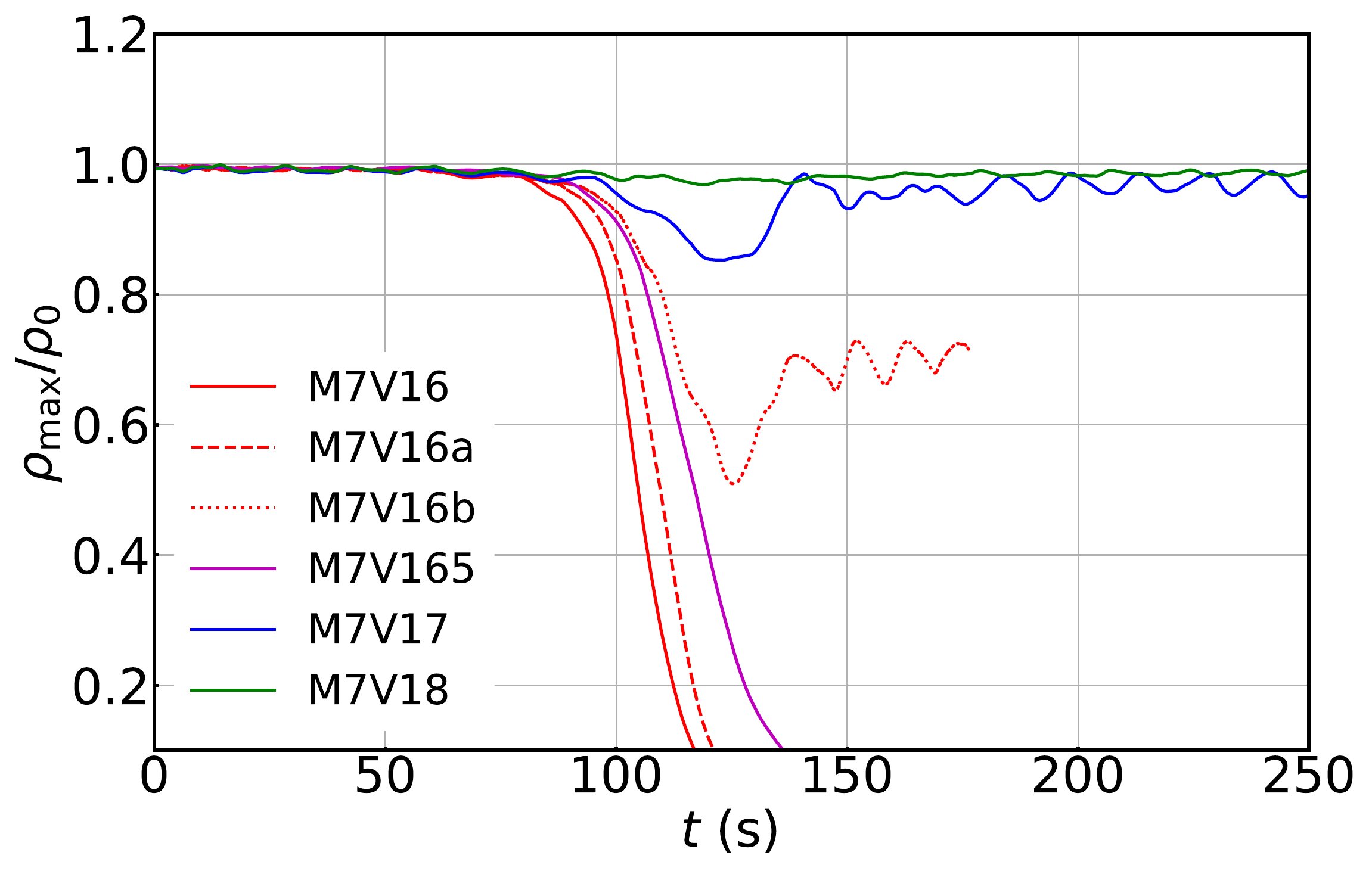}
  \caption{The maximum density as a function of time for $V=0.160$, 0.165, 0.170, and 0.180 with $M_*=0.7M_\odot$. The maximum density is normalized by the initial value denoted by $\rho_0$.
\label{rhomaxM7}}
\end{figure}
Figure~\ref{rhomaxM7} plots the evolution of the maximum density for $V=0.160$, 0.165, 0.170, and 0.180 with $M_*=0.7M_\odot$. We note that for $M_{\rm BH}=10^5M_\odot$, the orbital period for these parameters are in the range from $\approx 220$\,s for $V=0.160$ to $\approx 250$\,s for $V=0.180$.  The figure shows the results expected in the previous paragraphs: For \texttt{M7V16} and \texttt{M7V165}, the white dwarfs are tidally disrupted while approaching the black hole irrespective of the white-dwarf mass. For \texttt{M7V17} ($\beta=0.44$), the white dwarf is perturbed by the black hole near the periastron but it is not tidally disrupted. After the close encounter, the white dwarf is in an oscillating state due to the instantaneous tidal force received from the black hole.  
By contrast, for \texttt{M7V18}, the maximum density is approximately preserved to be constant, suggesting no disruption occurs and the tidal effect is negligible. Note that such tidal field may still perturb the white dwarf and produce detectable electromagnetic or gravitational-wave signal if a sufficient amplitude of oscillation is induced.

In Fig.~\ref{rhomaxM7}, the results of \texttt{M7V16} ($\beta=0.65$), \texttt{M7V16a} ($\beta=0.55$) and \texttt{M7V16b} ($\beta=0.45$) are also compared. As expected, for the first two models, the white dwarfs are tidally disrupted, while for the most compact white dwarf, the tidal disruption does not occur although it is perturbed significantly by the black-hole tidal force. This illustrates that the $\beta$ parameter is a good indicator for assessing whether tidal disruption takes place or not irrespective of the white-dwarf radius. 

\begin{figure}[t]
\includegraphics[width=88mm]{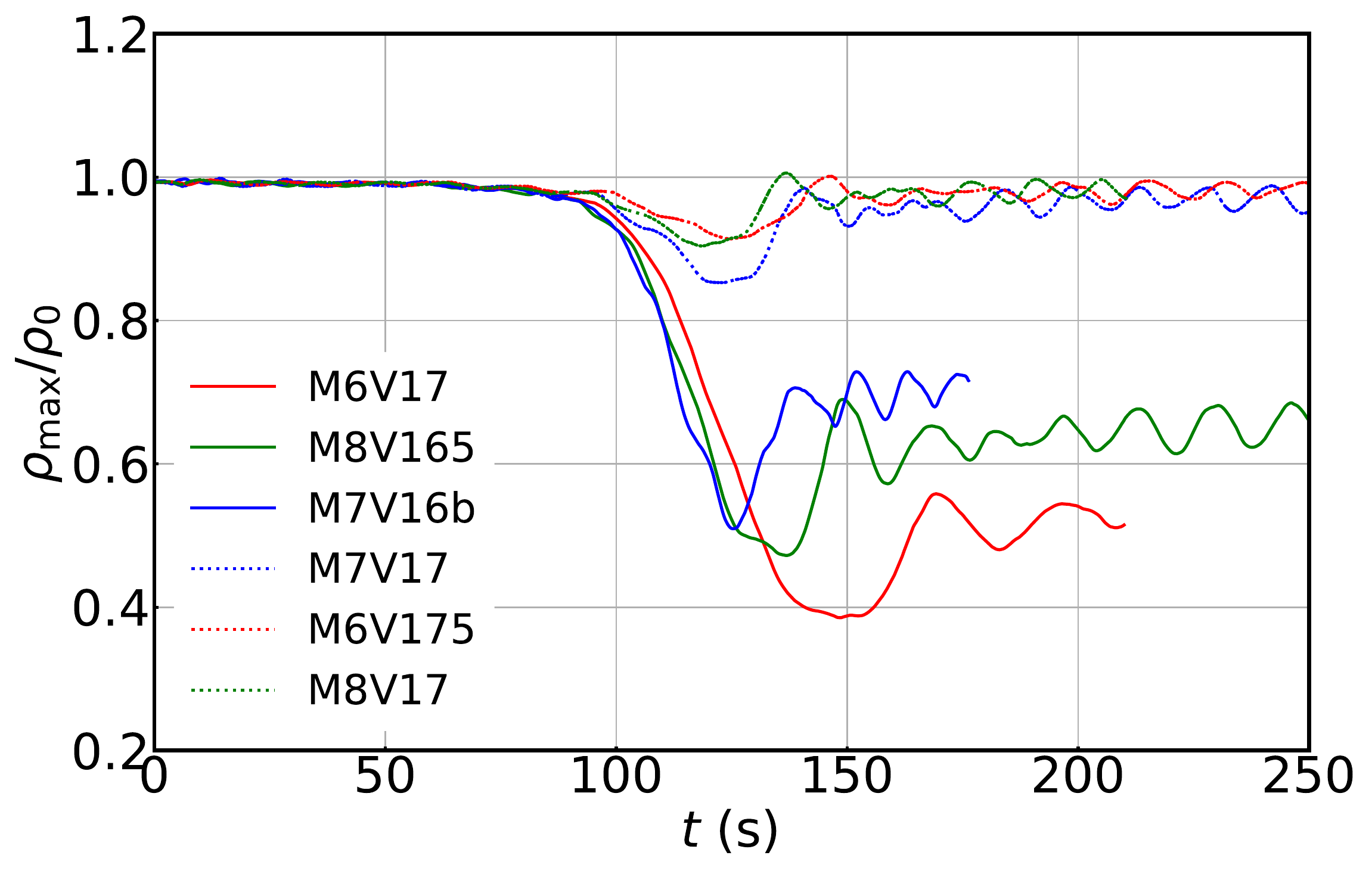}
  \caption{The maximum density as a function of time for the cases that stellar oscillation occurs $(0.5\agt \beta \agt 0.4)$. The red, blue, and green curves show the results with $M_*=0.6 M_\odot$, $0.7 M_\odot$ and $0.8 M_\odot$,  respectively. The maximum density is normalized by the initial value denoted by  $\rho_0$. 
\label{rhomaxOC}}
\end{figure}
Figure~\ref{rhomaxOC} shows the evolution of the maximum density when stellar oscillation is induced. For \texttt{M6V17} ($\beta=0.49$), the white dwarf is significantly elongated by the tidal force from the black hole; the central density is decreased to less than 50\% of the original value after passing through the periastron. Associated with the tidal effect, the mass is lost from the white dwarf. However, with the increase of the orbital radius, the central density increases again, resulting in a less massive white dwarf. This is also the case for \texttt{M8V165} ($\beta=0.47$) and \texttt{M7V16b} ($\beta=0.45$). 
These results indicate that the critical value of $\beta$ for the tidal disruption is $\sim 0.50$ and the threshold value for exciting a high-amplitude oscillation is $\beta\sim 0.45$. Figure~\ref{rhomaxOC} also shows that even for $0.40 \alt \beta \alt 0.45$ an appreciable oscillation is excited by the tidal force.

\begin{figure}[t]
\includegraphics[width=88mm]{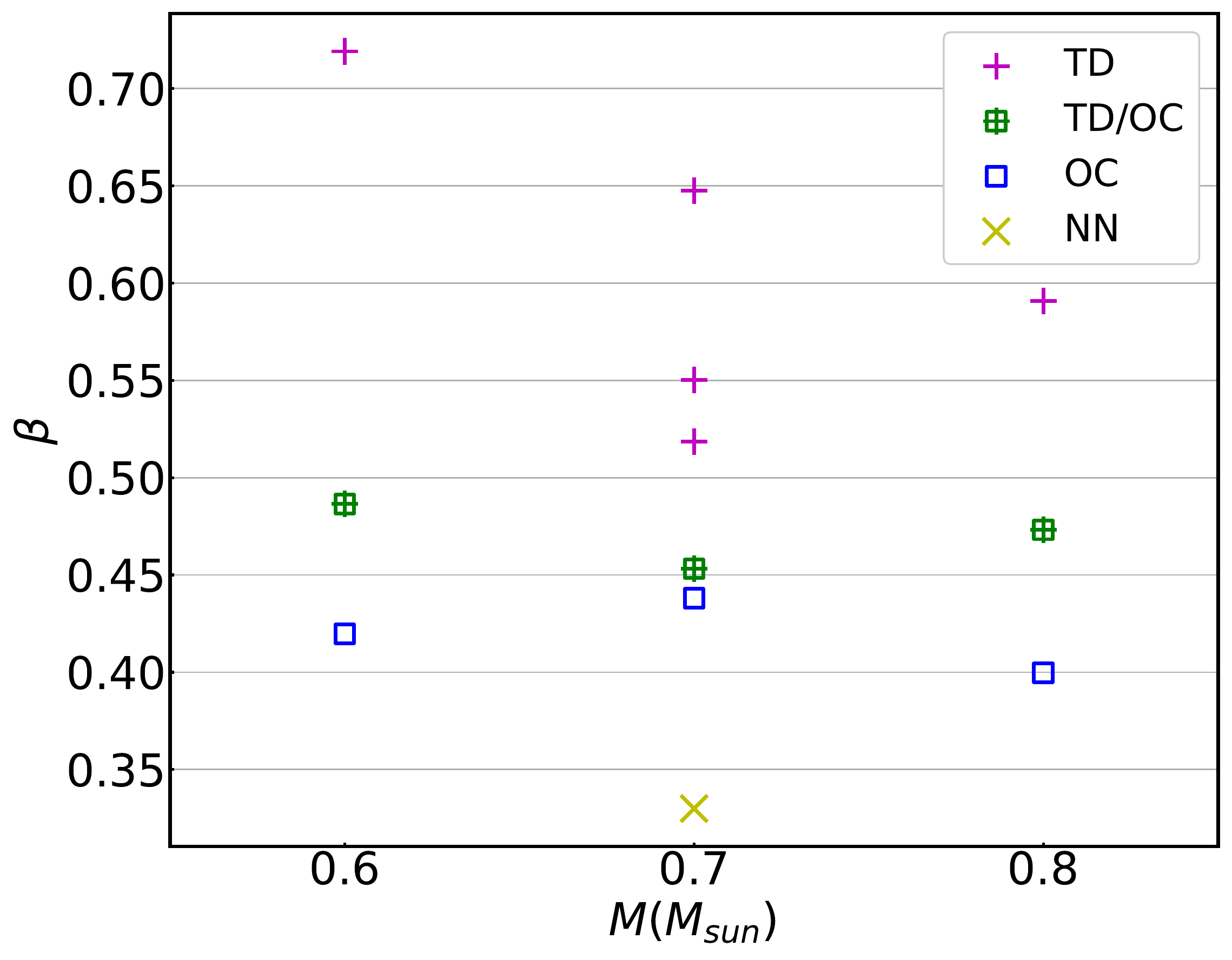}
  \caption{A summary for the fate of the white dwarfs in the plane of $M_*$ and $\beta$. TD and OC denote that tidal disruption and appreciable oscillation of the white dwarfs are observed after the close encounter of the white dwarfs with the black hole. NN denotes that no appreciable tidal effect is observed. 
\label{beta}}
\end{figure}

In Fig.~\ref{beta} and Table~\ref{table1}, we summarize the fates of white dwarfs as a result of the tidal interaction.  It is found that for $\beta \agt 0.5$ tidal disruption takes place and for $\beta \agt 0.4$, the white dwarfs are perturbed appreciably by the black-hole tidal field. All these results agree approximately with the expectation from the previous studies. 

\begin{figure*}[t]
\includegraphics[width=150mm]{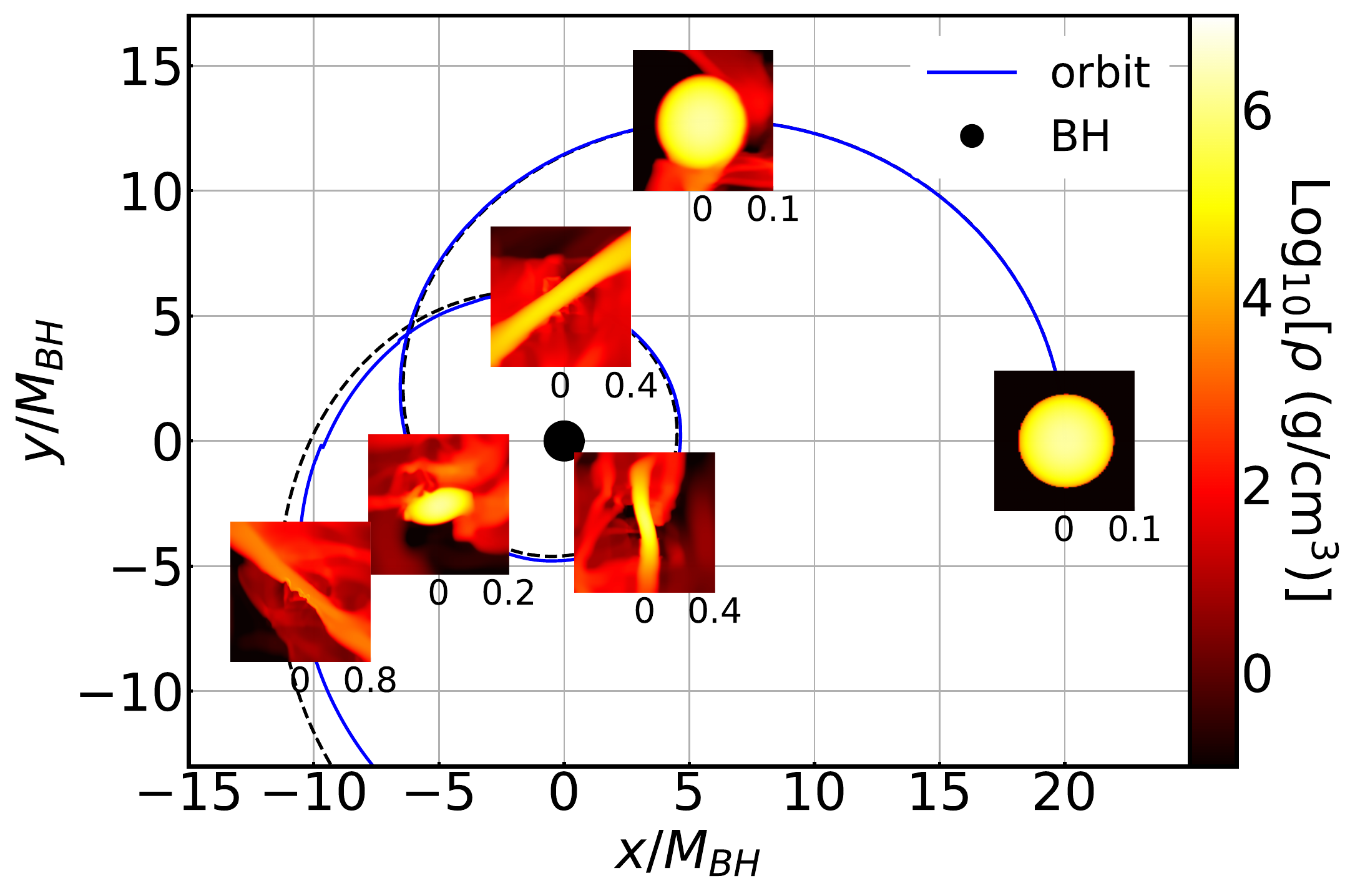}
  \caption{The density profiles of the tidally-disrupted white dwarf for the model $V=0.160$ and $M_*=0.7M_\odot$ (M7V16). The units of the length scale for the density plots are $GM_{\rm BH}/c^2 \approx 1.48 \times 10^5$\,km. The solid and dashed curves show the time evolution for the location of the maximum density and the elliptic orbit shown in Fig.~\ref{orbit} for $V=0.160$ (i.e., geodesic). The length scale of $x$ and $y$ axes is shown in units of $M_{\rm BH}$. 
\label{snapshotxy}}
\end{figure*}

For \texttt{M7V16}, tidal disruption takes place but only a small fraction of the white dwarf matter falls into the black hole because the fluid elements have specific angular momentum large enough to escape capturing by the black hole. Most of the tidally disrupted matter approximately maintains the original elliptic orbit (see Fig.~\ref{snapshotxy}) although the matter has an elongated profile. To clarify the eventual matter distribution around the black hole, we will need to follow the matter motion for more than 10 orbits. This topic is one of our major research targets in the future.

For $0.4 \alt \beta \alt 0.5$, the white dwarf will be continuously perturbed by the black-hole tidal force whenever it passes through the periastron.  In addition the angular momentum is transported during the tidal interaction, and it will lead to the transport of the orbital angular momentum to the white dwarf resulting in a spin-up of it.  According to a perturbation study for the stellar encounter, the energy deposition during the tidal interaction in one orbit is written approximately as~\cite{PT77}
\beqn
\Delta E_{\rm tid}&=&f_{\rm tid}\left(\frac{M_*^2}{R_*}\right)
\left(\frac{M_{\rm BH}}{M_*}\right)^2
\left(\frac{R_*}{r_{p,A}}\right)^6 \nonumber \\
&=&f_{\rm tid}\left(\frac{M_*^2}{R_*}\right)\beta^6. 
\eeqn
where $f_{\rm tid}$ is a factor of $O(0.1)$, which depends on $\beta$ and the equation of state. Associated with the energy deposition near the periastron, the angular momentum deposition is also deposited. In one orbit it is approximately estimated by $\Delta J_{\rm spin} \approx \Delta E_{\rm tid}/\Omega_p$ ~\cite{Kumar1998} where $\Omega_p=\sqrt{M_{\rm BH}/r_{p,A}^3}$, and thus,
\beqn
\Delta J_{\rm spin}&=&f_{\rm spin} M_* \sqrt{M_*R_*}
\left(\frac{M_{\rm BH}}{M_*}\right)^{3/2}
\left(\frac{R_*}{r_{p,A}}\right)^{9/2}\nonumber \\
&=&f_{\rm spin} M_* \sqrt{M_*R_*} \beta^{9/2}, \label{Jtid}
\eeqn
where $f_{\rm spin}$ is a coefficient of the same order of the magnitude of $f_{\rm tid}$. 
\begin{figure}[t]
\includegraphics[width=88mm]{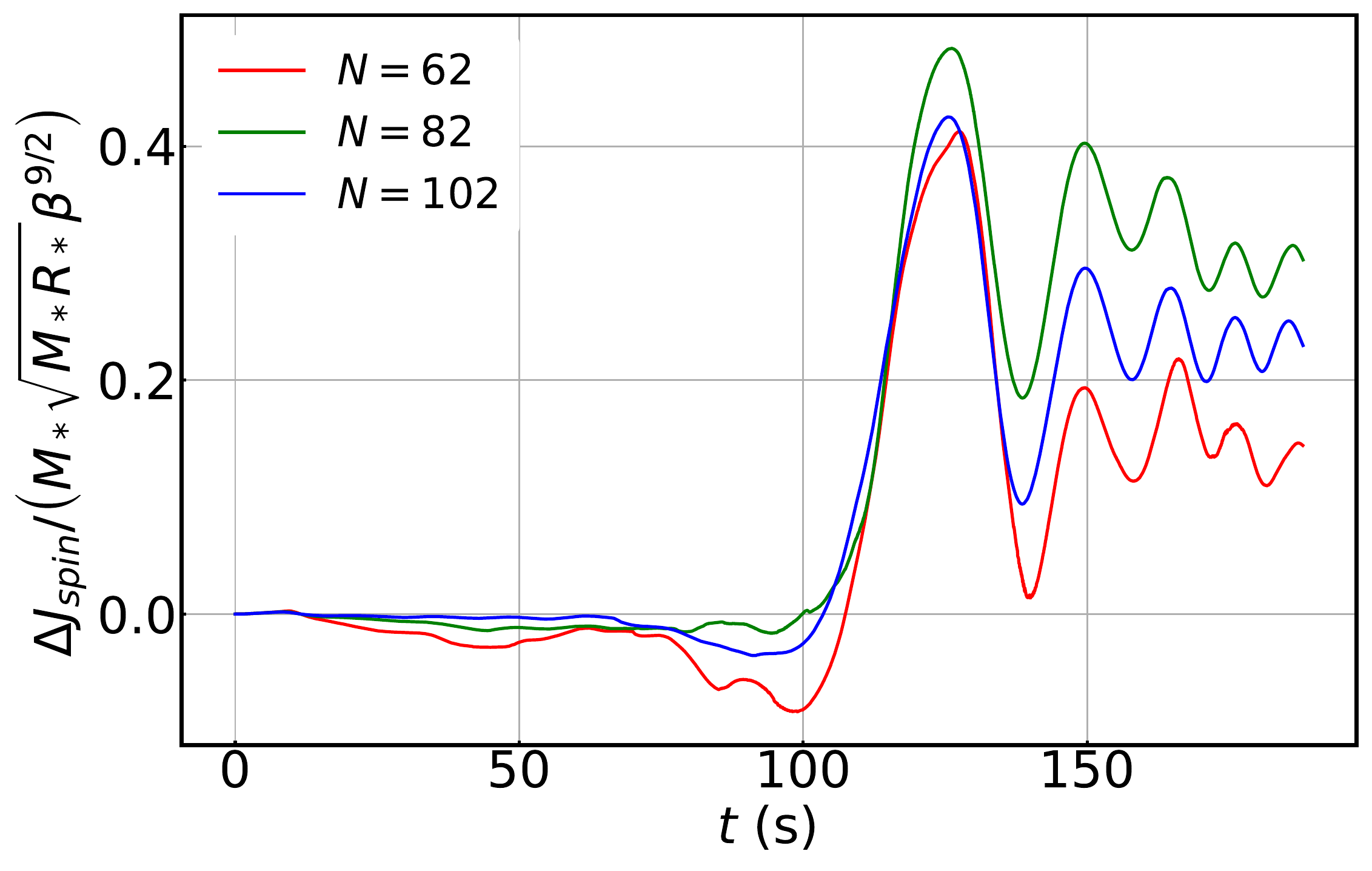}
  \caption{The rescaled change in angular momentum
  $\Delta J_{spin} / \left( M_* \sqrt{M_*R_*} \beta^{9/2} \right)$
  as a function of time for the stellar oscillation scenario (\texttt{M8V17}).
  This agrees with the analytic expression Eq.~(\ref{Jtid}) if $f_{\rm spin} \sim 0.1$--0.3.
\label{spin}}
\end{figure}

Because the maximum spin angular momentum of the star is approximately written as $M_* \sqrt{M_*R_*}$, we find that $\Delta J_{\rm spin}$ can be more than 0.1\% of the maximum spin if a white dwarf passes through a close orbit with $\beta \agt 0.4$. We approximate the orbit angular momentum $J_{\rm orbit}$ and the spin angular momentum $J_{\rm spin}$ of the white dwarf as
\beqn
J_{\rm orbit} =&& M_h \left(\left<x\right>\left<u_y \right> 
                          - \left<y\right>\left<u_x \right>\right), \\
J_{\rm spin} =&& \int d^3x \psi^6 \rho h \left[\left(x - \left<x\right>\right)\left(u_y - \left<u_y\right>\right)\right. \nonumber \\
                            && \left.-\left(y - \left<y\right>\right)\left(u_x - \left<u_x\right>\right)\right],
\eeqn
where $M_h:=\int d^3 x \psi^6 \rho h$ and the volume average of quantity $q$ is defined as $\left<q\right>:=\frac{1}{M_h}\int d^3 x \psi^6 \rho h q$. In such decomposition, the sum of orbital and spin angular momentum equals to the total angular momentum of the white dwarfs.
We analyzed the spin angular momentum gain of the white dwarfs for \texttt{M8V17}, and we indeed find $\Delta J_{\rm spin}/(M_*\sqrt{M_*R_*}\beta^{9/2}) \approx 0.1$--0.3 as shown in Fig.~\ref{spin}. %
Note that the spin up of white dwarf $\Delta J_{\rm spin}$ is about $10^{-6}$ of the total angular momentum, and  hence, it is not easy to determine $\Delta J_{\rm spin}$ accurately.
Although we cannot achieve a good convergence in $\Delta J_{\rm spin}$, 
we are able to obtain a noticeable rise in $J_{\rm spin}$ during the close encounter, which suggests $f_{\rm spin} \sim 0.1$--0.3, which is consistent with the above analytic result.

For close orbits, the tidal angular-momentum transport can dominate over the orbital angular momentum loss by gravitational-radiation reaction.  Assuming that gravitational waves are most efficiently emitted near the periastron at which we may approximate the orbit to be circular, the angular momentum dissipation by gravitational waves in one orbit can be written as~\cite{Peters}
\beqn
\Delta J_{\rm GW} \approx \frac{64 \pi}{5}\frac{M_{\rm BH}^2
  M_*^2}{r_{p,A}^2}\left(1 + \frac{7e^2}{8}\right),
\eeqn
where $e$ denotes the eccentricity. Thus, the ratio of
$\Delta J_{\rm tid}$ to $\Delta J_{\rm GW}$ is written as
\beqn
    \frac{\Delta J_{\rm spin}}{\Delta J_{\rm GW}}&\approx& 23f_{\rm spin}
    \left(\frac{r_{p,A}}{4M_{\rm BH}}\right)^{-5/2}
    \left(\frac{M_{\rm BH}}{10^5M_\odot}\right)^{-3}
    \left(\frac{M_*}{0.7M_\odot}\right)^{-2}
    \nonumber \\
    && \times \left(\frac{R_*}{10^4\,{\rm km}}\right)^5
    \left(1 + \frac{7e^2}{8}\right)^{-1}.
\eeqn
Thus it is larger than unity for $r_{p,A} \alt 7M_{\rm BH}/c^2$, $R_* \approx 10^4$\,km, $M_{\rm BH}=10^5M_\odot$, $M_*=0.7M_\odot$, and $f_{\rm spin}=0.2$.  This is also the case for the ratio of $\Delta E_{\rm tid}/\Delta E_{\rm GW}$ where $\Delta E_{\rm GW}$ is the energy dissipated by gravitational waves in one orbit.  Thus, near the tidal disruption orbit, the orbital evolution would be primarily determined not by the gravitational-wave emission but by the tidal effect. To clarify the eventual fate of such a white dwarf, we obviously need a long-term accurate simulation. Such a topic is one of our future targets.

We note that both $\Delta J_{\rm spin}$ and $\Delta J_{\rm GW}$ are much smaller than the orbital angular momentum of order $M_* \sqrt{M_{\rm BH} r_{p,A}}$. Thus, the cumulative effect of the tidal angular momentum transport plays an important role just prior to the tidal disruption. By repeated tidal interaction, the spin angular velocity of the white dwarfs is likely to be enhanced up to $\sim M_{\rm BH}^{1/2}/r_p^{3/2}=\beta^{3/2}M_*^{1/2}/R_*^{3/2}$. In addition, the stellar oscillation for which the oscillation energy is comparable to or larger than the rotational kinetic energy should be excited. As a result, mass loss could be induced, resulting in the increase of the stellar radius and enhancing the importance of tidal interaction. 
In this type of the system, the tidal disruption is unlikely to take place by one strong impact by the black-hole tidal force but likely to do as a result of a secular increase of the stellar radius (see, e.g., Refs.~\cite{Dai:2011wyd,Linial:2022bjg} for related studies).


\section{Summary}\label{secIV}

We reported a new numerical-relativity code which enables us to explore tidal disruption of white dwarfs by a relatively low-mass supermassive black hole. As a first step toward more detailed future studies, we paid attention to the condition for tidal disruption of white dwarfs with typical mass range in elliptic orbits by a nonspinning supermassive black hole. We showed that our code is capable of determining the condition for the tidal disruption. As expected from previous general relativistic works (e.g., Refs.~\cite{Ishii,Ryu4}), the tidal disruption takes place for $\beta \agt 0.5$ and an appreciable oscillation of the white dwarfs are induced by the black-hole tidal effect for $\beta \agt 0.4$ for orbits close to the black hole in the $\Gamma=5/3$ polytropic equation of state. The critical value for the onset of the tidal disruption is smaller than that obtained by Newtonian analysis.  For white dwarfs with $M_*=0.6M_\odot$ and $R_*=1.2\times 10^4$\,km, $\beta$ can be larger than 0.4 even for $M_{\rm BH} \approx 4 \times 10^5M_\odot$ if the periastron radius is $r_{p,A}=4M_{\rm BH}$. Our result indicates that in such systems with a relatively low-mass (but not intermediate-mass) supermassive black hole for which gravitational waves in the late inspiral phase can be detected by LISA~\cite{LISA}, tidal disruption can occur for typical white dwarfs. For spinning black holes with the dimensionless spin parameter of $\agt 0.9$, $r_{p,A}$ can be smaller than $\sim 1.7M_{\rm BH}$~\cite{BPT72}. For such black holes, tidal disruption of typical-mass white dwarfs may occur even for $M_{\rm BH} \approx 10^6M_\odot$. Investigation of this possibility is a future issue.


There are several issues to be explored. The first one is to extend our implementation for spinning black holes. Since no analytic solution is known for the spacetime of spinning black holes on the limit hypersurface, we need to develop a method to provide $g_{\mu\nu}^0$ for employing the formulation introduced in this paper. One straightforward way to prepare such data is just to numerically perform a simulation for a spinning black hole (in vacuum) until the hypersurface reaches the limit hypersurface as a first step, and then, the obtained data are saved and used in the subsequent simulations with white dwarfs. A more subtle issue along this line is to prepare the initial condition. For nonspinning black holes, we can assume that the conformal flatness of the three metric, and as a result, the initial-value equations are composed only of elliptic-type equations with the flat Laplacian. For the spinning black holes, the basic equations are composed of elliptic-type equations of complicated Laplacian, and hence, the numerical computation could be more demanding, although in principle it would be still possible to obtain an initial condition. We plan to explore this strategy in the subsequent work.

For modeling realistic white dwarfs it is necessary to implement a realistic equation of state. If we assume that the temperature of the white dwarfs is sufficiently low and the pressure is dominated by that of degenerate electrons, it is straightforward to implement this.

More challenging issue is to follow the hydrodynamics of tidally disrupted white-dwarf matter for a long term. After the tidal disruption, the matter of the white dwarf is likely to move around the black hole for many orbits.  During such orbits, the matter collides each other, and eventually, a compact disk will be formed after the circularization. Such disks are likely to be hot due to the shock heating, and thus, it can be a source of electromagnetic counterparts of the tidal disruption.  In the presence of magnetic fields, magnetorotational instability~\cite{BH98} occurs in the disk, and the magnetic fields will be amplified. If the amplified magnetic field eventually penetrates the black hole and if the black hole is appreciably spinning, a jet may be launched through the Blandford-Znajek effect~\cite{BZ77}. After the amplification of the magnetic fields, a turbulent state will be developed in the disk and mass ejection could occur by the effective viscosity or magneto-centrifugal force~\cite{BP82}. The ejecta may be a source of electromagnetic signals. One long-term issue is to investigate such scenarios by general relativistic magnetohydrodynamics.

\acknowledgments

We thank Kenta Hotokezaka for helpful discussion. This work was in part supported by Grant-in-Aid for Scientific Research (Grant No.~JP20H00158) of Japanese MEXT/JSPS.  Numerical computations were in part performed on Sakura clusters at Max Planck Computing and Data Facility.

\bibliographystyle{apsrev4-2}
\bibliography{refs}

\end{document}